\documentclass[12pt]{article}
\usepackage{graphicx}

\newcommand\pubnumber{}
\newcommand\pubdate{\today}

\def\glasgow{School of Physics and Astronomy\\
University of Glasgow, Glasgow G12 8QQ, UK}
\def\support{\footnote{Speaker}}

\def\Title#1{\begin{center} {\Large #1 } \end{center}}
\def\Author#1{\begin{center}{ \sc #1} \end{center}}
\def\Address#1{\begin{center}{ \it #1} \end{center}}

\newcommand\pubblock{\rightline{\begin{tabular}{l} \pubnumber\\
         \pubdate  \end{tabular}}}
\newenvironment{Abstract}{\begin{quotation}  }{\end{quotation}}
\newenvironment{Presented}{\begin{quotation} \begin{center} 
             PRESENTED AT\end{center}\bigskip 
      \begin{center}\begin{large}}{\end{large}\end{center} \end{quotation}}
\def\Acknowledgements{\bigskip  \bigskip \begin{center} \begin{large}
             \bf ACKNOWLEDGEMENTS \end{large}\end{center}}

\def\beq{\begin{equation}}
\def\eeq#1{\label{#1}\end{equation}}
\def\eeqn{\end{equation}}

\def\beqa{\begin{eqnarray}}
\def\eeqa#1{\label{#1}\end{eqnarray}}
\def\eeqan{\end{eqnarray}}
\def\CR{\nonumber \\ }

\let\bar=\overbar

\def\Dslash{\not{\hbox{\kern-4pt $D$}}}
\def\dslash{\not{\hbox{\kern-2pt $\del$}}}

\def\msb{{\bar{\ssstyle M \kern -1pt S}}}

\begin{document}
\begin{titlepage}
\pubblock

\vfill
\Title{$D\to K\ell\nu$ and $D\to \pi\ell\nu$ form factors from Lattice QCD}
\vfill
\Author{ J. Koponen\support, C.T.H. Davies, G. Donald}
\Author{ HPQCD Collaboration}
\Address{\glasgow}
\vfill
\begin{Abstract}
We present a very high statistics study of $D$ and $D_s$
semileptonic decay form factors on the lattice. We work
with MILC $N_f=2+1$ lattices and use the Highly Improved
Staggered Quark action (HISQ) for both the charm and the
strange and light valence quarks. We use both scalar and
vector currents to determine the form factors $f_0(q^2)$
and $f_+(q^2)$ for a range of $D$ and $D_s$ semileptonic
decays, including $D\to \pi\ell\nu$ and $D\to K\ell\nu$. By using
a phased boundary condition we are able to tune  accurately
to $q^2=0$ and explore the whole $q^2$ range allowed by
kinematics. We can thus compare the shape in $q^2$ to that
from experiment and extract the CKM matrix element $|V_{cs}|$.
We show that the form factors are insensitive to the
spectator quark: $D \to K\ell\nu$ and $D_s \to \eta_s\ell\nu$
form factors are essentially the same, which is also true
for $D\to \pi\ell\nu$ and $D_s\to K\ell\nu$ within 5\%. This
has important implications when considering the corresponding
$B/B_s$ processes.
\end{Abstract}
\vfill
\begin{Presented}
The 5th International Workshop on\\ Charm Physics (Charm 2012)\\
Honolulu, Hawai'i, May 14--17, 2012
\end{Presented}
\vfill
\end{titlepage}
\def\thefootnote{\fnsymbol{footnote}}
\setcounter{footnote}{0}

\section{Introduction}

Lattice QCD is an excellent tool for calculating strong interaction
effects from first principles and can provide accurate phenomenology
not available with any other method. Particularly important and
interesting applications of this method are the determinations of
various heavy meson weak decay matrix elements that are key to
constraining the vertex of the Unitarity Triangle derived from
the Cabibbo-Kobayashi-Maskawa (CKM) matrix. This provides  a
stringent test of the self-con\-sis\-ten\-cy of the Standard Model.

However, calculations of these matrix elements must not be seen in
isolation. This is only one part of the HPQCD program to calculate
meson spectra, decay constants and other QCD observables fully
non-perturbatively (see e.g. the calculation of $f_{D_s}$
in~\cite{Dsdecayconst} or $J/\psi$ mass, leptonic width and
radiative decay rate to $\eta_c$ in~\cite{Jpsietac}). The Highly
Improved Staggered Quark (HISQ) formalism enables us to keep
the discretization errors small and to treat charm quarks the
same way as light and strange quarks, which reduces the
systematic errors.

A simple recipe for doing a Lattice QCD calculation is:
\begin{enumerate}
\item Generate sets of gluon fields from Monte Carlo integration
of the QCD path integral (including effects of u, d and s sea quarks)

\item Calculate averaged “hadron correlators” from valence quark
propagators

\item Fit the correlators as a function of time to obtain masses
and matrix elements

\item Determine a and fix $m_q$ to get results in physical units

\item Extrapolate to $a=0$ and physical light quark mass for
real world
\end{enumerate}
However, in this paper we don't go into details of Lattice QCD
and concentrate on the results. The paper is organized as
follows: In Section~\ref{sect:lattFF} we briefly explain how we
calculate the form factors on the lattice and fit the results.
We give our results for the form factors in
Section~\ref{sect:semileptonicFF} and talk about the
$z$-expansion and continuum and chiral extrapolations in
more detail in Section~\ref{sect:zexpansion}. In
Section~\ref{sect:extractVcs} we explain how we extract $V_{cs}$
and give our result and summarise in Section~\ref{sect:summary}.

\section{Form factors on the lattice} 
\label{sect:lattFF}

\begin{figure}[htb]
\begin{center}
\includegraphics[width=0.33\textwidth]{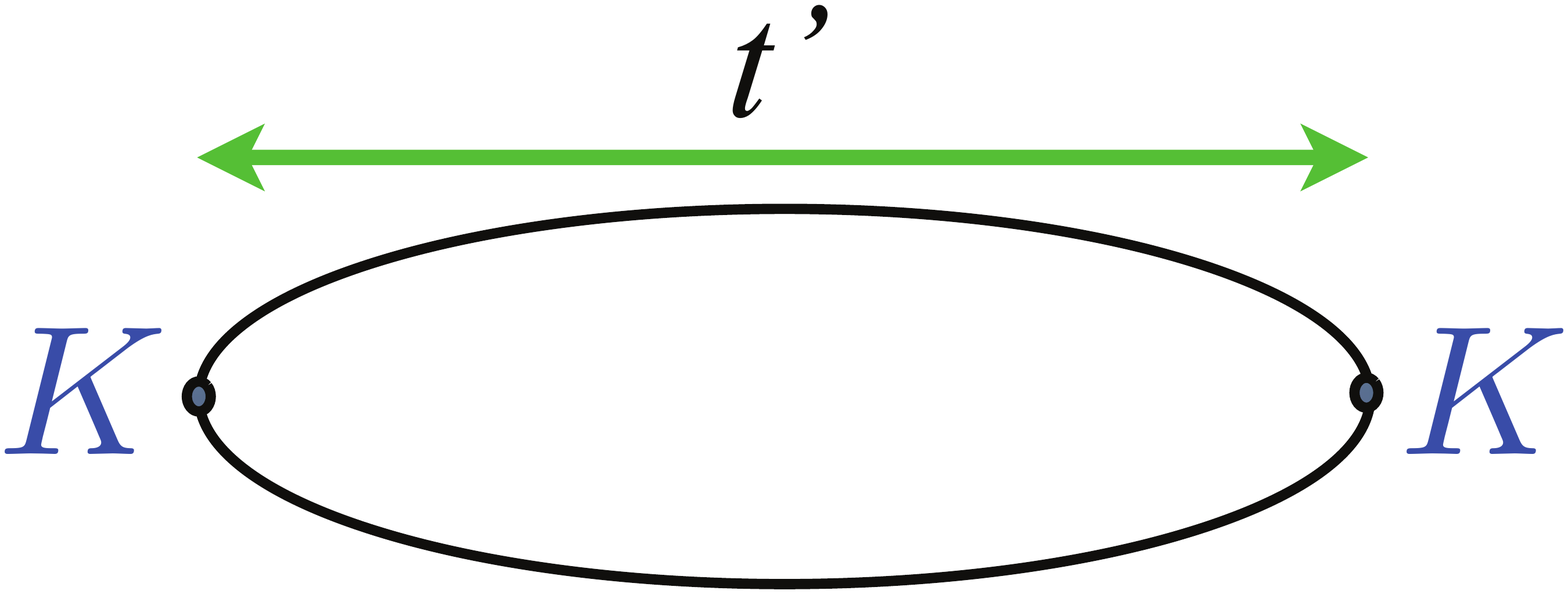}
\includegraphics[width=0.32\textwidth]{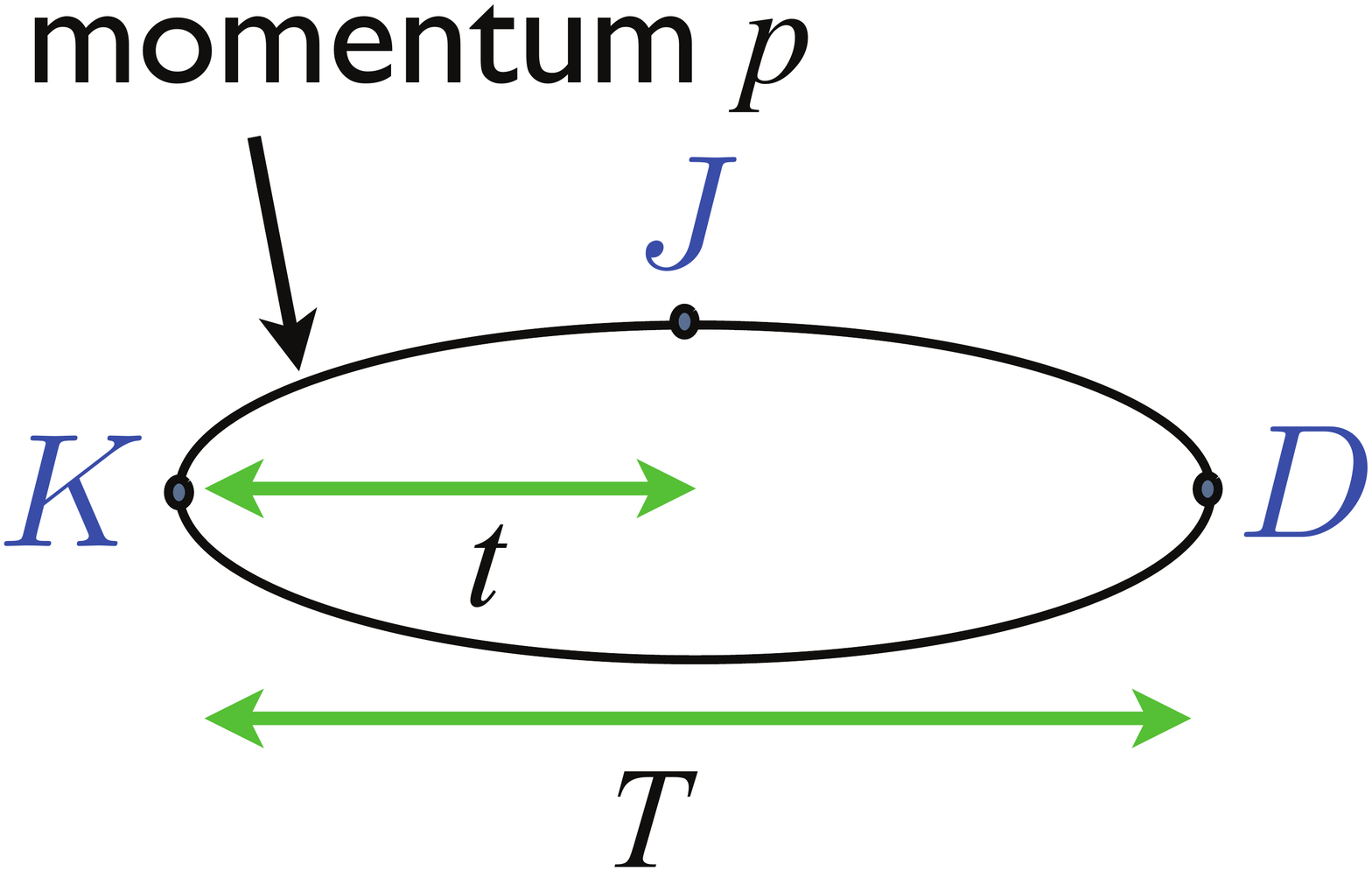}
\end{center}
\caption{2-point and 3-point correlators. In the 2-point correlator
a kaon is created at time $t'$ and annihilated at time 0. In
the 3-point correlator the two mesons are time $T$
apart and a scalar or vector current $J$ is inserted at time $t$ ($0<t<T$).
One of the quarks can be given a momentum $p$ to explore the
full $q^2$ range.}
\label{fig:2ptn3ptcorr}
\end{figure}

Semileptonic form factors are 3-point amplitudes in Lattice QCD.
To get information of both form factors, $f_0(q^2)$ and $f_+(q^2)$, we
calculate scalar and vector currents and the corresponding 2-point
correlators for the mesons -- see Fig.~\ref{fig:2ptn3ptcorr}.
The scalar current is
\begin{equation}
\langle K|S|D\rangle = f_0^{D \to K}(q^2)\frac{M_D^2-M_K^2}{m_{0c}-m_{0s}},
\end{equation}
where $q^\mu=p_D^\mu-p_K^\mu$ (using $D\to K\ell\nu$ as an example), and
the vector current can be written as
\begin{equation}
\langle K|V^\mu|D\rangle = f_+^{D \to K}(q^2)\Bigg[p^\mu_D+p^\mu_K
-\frac{M_D^2-M_K^2}{q^2}q^\mu\Bigg]+f_0^{D \to K}(q^2)
\frac{M_D^2-M_K^2}{q^2}q^\mu.
\end{equation}
Note that this guarantees that $f_0(0)=f_+(0)$. See also our earlier
calculation of semileptonic form factors in~\cite{DKsemileptonic}.
We use 3 ensembles of MILC asqtad $N_f=2+1$ lattice configurations
(sets 1, 2 and 4 in~\cite{Jpsietac}).

From experiments we get the differential decay rates, e.g.
\begin{equation}
\label{eq:diffdecayrate}
\frac{d\Gamma}{dq^2} = \frac{G_F^2p_K^3}{24\pi^3}
|V_{cs}|^2|f_+^{D \to K}(q^2)|^2
\end{equation}
for $D\to K\ell\nu$. From this one can determine
$|V_{cs}\cdot f_+^{D \to K}(q^2)|$, but one needs either $f_+(q^2)$
from theory or $V_{cs}$ from unitarity to determine the other.

\subsection{Fitting the Lattice QCD results} 

We fit the lattice 2-point and 3-point correlators simultaneously,
as a function of time $t'$ and meson separation $T$ (see
Fig.~\ref{fig:2ptn3ptcorr}) to estimate correlations between all
fit parameters. We use multi-exponential fits (up to 5 exponentials)
to reduce systematic errors from the excited states. The fit
parameters are constrained by Bayesian priors. More details about
the type of fits that we use can be found in~\cite{DKsemileptonic}.

We have very high statistics, of the order of 100000 correlators,
which gives us very small statistical errors. By giving one of
the quarks a momentum $p$ using twisted boundary
conditions~\cite{ETMC}, as shown in Fig.~\ref{fig:2ptn3ptcorr},
we can tune accurately to $q^2=0$ or choose any $q^2$ value in
the allowed kinematical region. This enables us to study the
whole physical $q^2$ range.

\section{Semileptonic form factors}
\label{sect:semileptonicFF}

We have calculated the form factors $f_0$ and $f_+$, as a function
of $q^2$, for various $D$ and $D_s$ semileptonic decays. Let us
divide the results into two groups: charm to strange decays and
charm to light decays.

\subsection{Charm to strange decay: $D\to K\ell\nu$,
$D_s\to \eta_s\ell\nu$ and $D_s\to \phi\ell\nu$}

\begin{figure}[htb]
\centering
\includegraphics[width=0.33\textwidth]{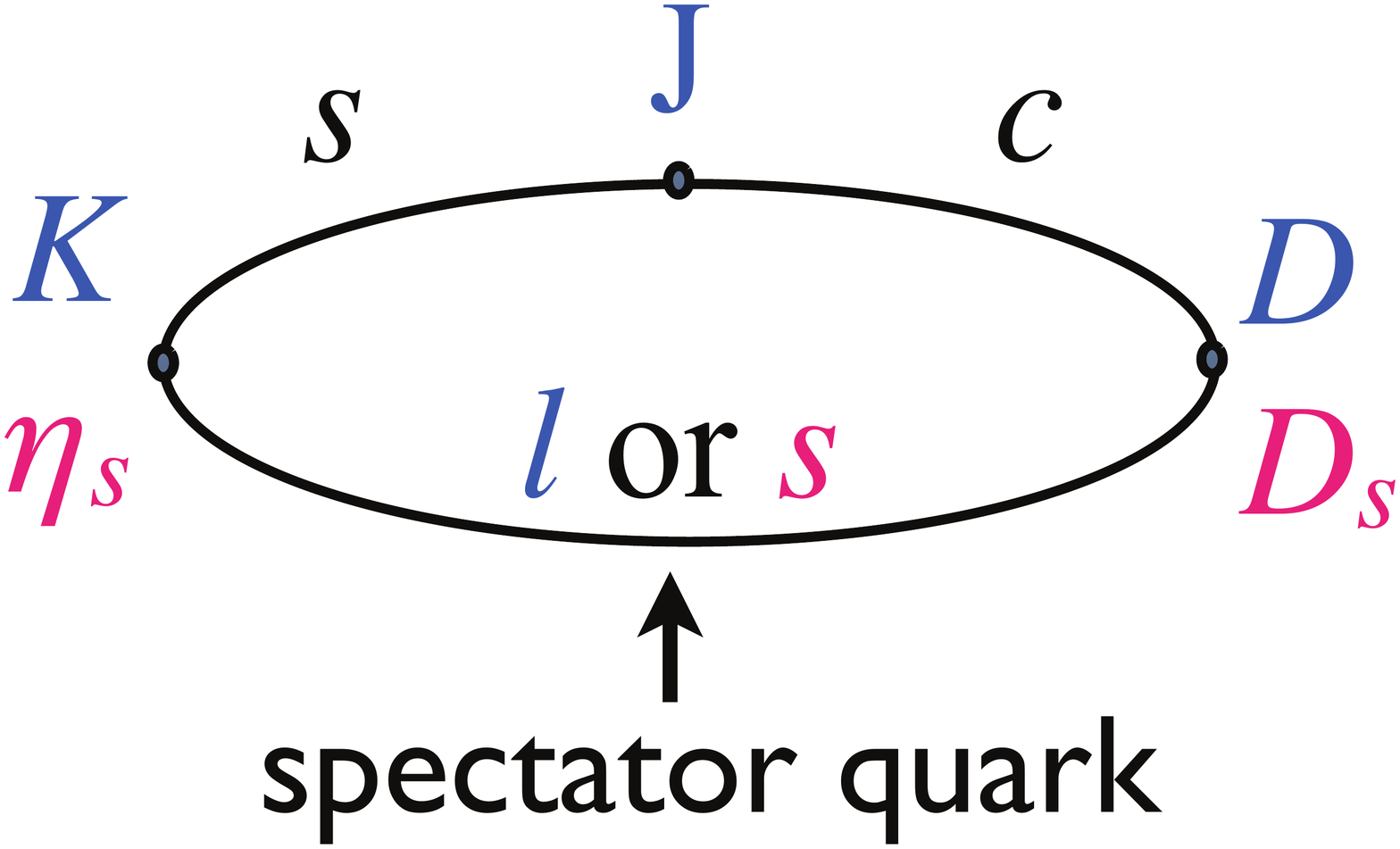}
\includegraphics[width=0.35\textwidth]{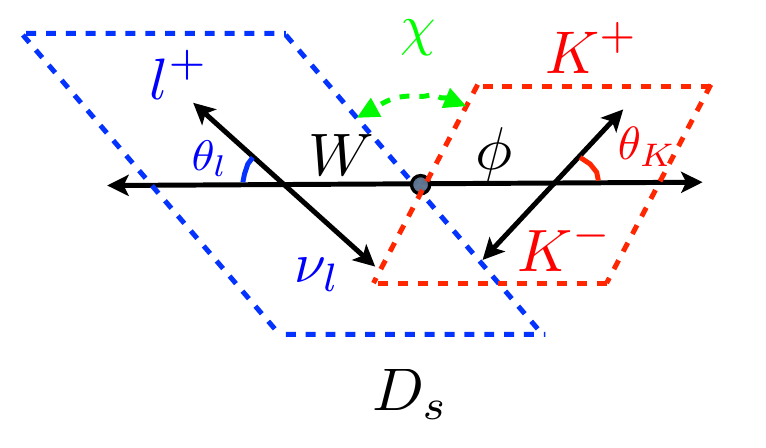}
\caption{On the left: $D\to K\ell\nu$ and $D_s\to \eta_s\ell\nu$
decays are the same except for the spectator quark; On the right:
Kinematics of the $D_s\to \phi\ell\nu$ decay.}
\label{fig:spectatordiagram}
\end{figure}

\begin{figure}[htb]
\centering
\includegraphics[width=0.75\textwidth]{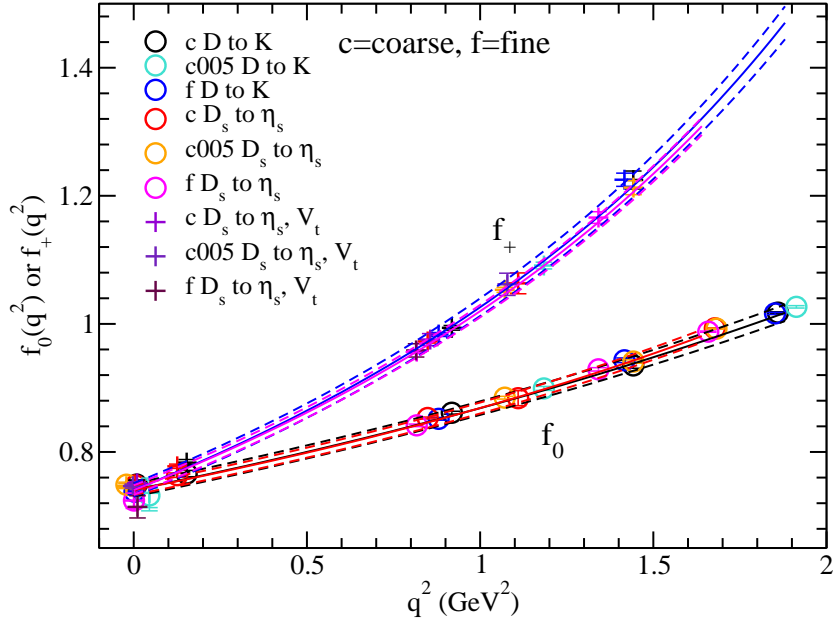}
\caption{$D\to K\ell\nu$ and $D_s\to \eta_s\ell\nu$ form factors.
The form factors are very insensitive to the spectator quark.}
\label{fig:DKnDsetas}
\end{figure}

\begin{figure}[htb]
\centering
\includegraphics[angle=-90,width=0.82\textwidth]{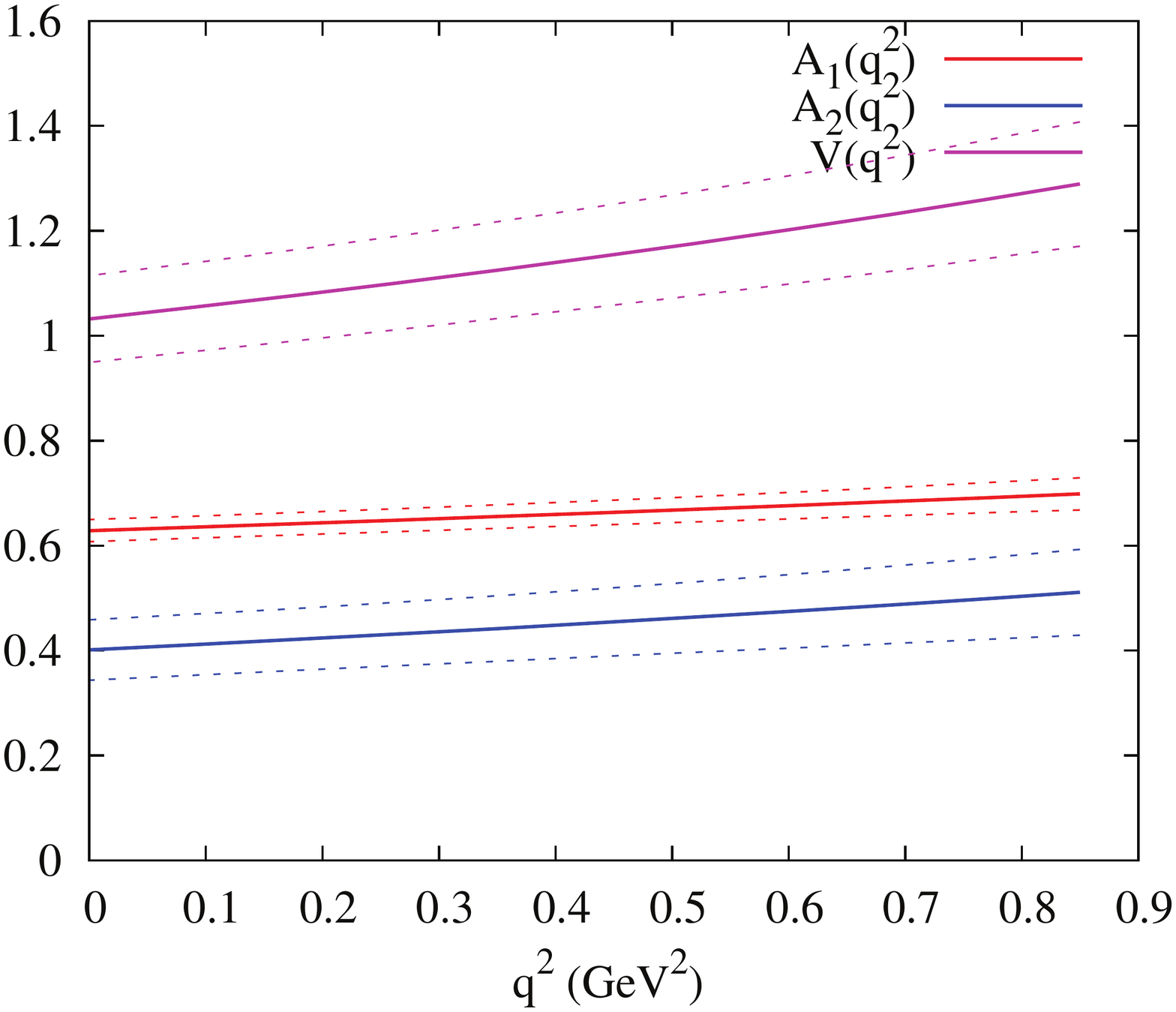}
\caption{$D_s\to\phi\ell\nu$ form factors, defined
in~\cite{Lattice2011Gordon}.}
\label{fig:DsphiFF}
\end{figure}

\begin{figure}[htb]
\centering
\includegraphics[angle=-90,width=0.85\textwidth]{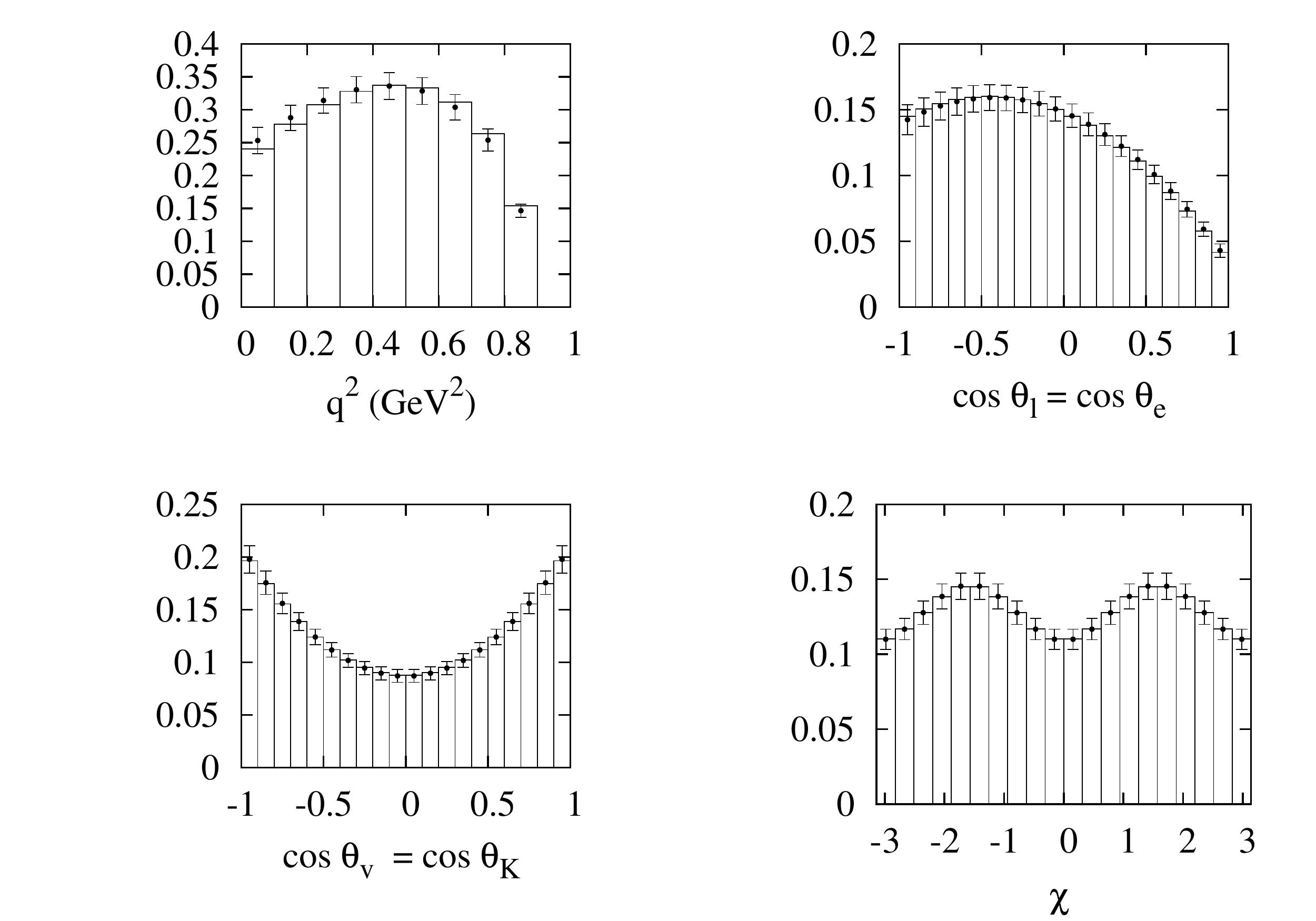}
\caption{$D_s\to\phi\ell\nu$ decay rates in $q^2$ bins and angular
distributions. The angles are defined in Fig.~\ref{fig:spectatordiagram}.}
\label{fig:Dsphi}
\end{figure}

$D\to K\ell\nu$ and $D_s\to \eta_s\ell\nu$ both have the same
charm-strange current, as they are both charm to strange decays.
$\eta_s$ is the pseudoscalar $s\bar{s}$ meson, so both decays
are pseudoscalar to pseudoscalar decays. Note that $\eta_s$ is
not a physical meson, but can be easily calculated on a lattice. 
The difference between these two decays is the spectator quark,
light vs. strange, as illustrated in Fig.~\ref{fig:spectatordiagram}.

Our results on different lattice ensembles are shown in
Fig.~\ref{fig:DKnDsetas}, along with the final result after
the continuum and chiral extrapolation. The $q^2$ dependence is
well understood qualitatively -- $f_+$ rises more steeply than
$f_0$, as $f_+$ is governed by the vector meson $M_{D^\ast_s}$ in
the pole mass parameterization, whereas $f_0$ is governed
by the scalar meson $M_{D^\ast_{s0}}$. The continuum and chiral
extrapolation will be discussed in more detail in
Section~\ref{sect:zexpansion}. The differences between the coarse
and the fine lattice results are small, i.e. the discretisation
effects are very well under control.

Note that the shapes of the form factors do not depend
on the spectator quark: the form factors for these two decays,
$D\to K\ell\nu$ and $D_s\to \eta_s\ell\nu$, are the same within 3\%
and even closer when one moves away from $q^2=0$.
Comparing the decay constants of the two mesons, $f_D$ and $f_{D_s}$,
one might expect a change of about 15\% when going from a light
spectator quark to a strange quark. However, this does not appear
to be the case for form factors.

On the lattice one can also calculate form factors for a pseudoscalar
to a vector meson semileptonic decay. $D_s\to \phi\ell\nu$ is a charm
to strange decay like $D\to K\ell\nu$ and $D_s\to
\eta_s\ell\nu$. As $\phi$ is a vector meson, there are now more
form factors than in the pseudoscalar to pseudoscalar decay. The
general form for a pseudoscalar to vector matrix element is
\beqa
\langle\phi (p',\epsilon )|V^\mu -A^\mu |D_s(p)\rangle =
\frac{2i\epsilon_{\mu\nu\alpha\beta}}{M_{D_s}+M_\phi}
\epsilon^\nu p^\alpha_{D_s}p^\beta_\phi V(q^2)
-(M_{D_s}+M_\phi)\epsilon^\mu A_1(q^2)\CR
+\frac{\epsilon\cdot q}{M_{D_s}+M_\phi}(p+p')A_2(q^2)
+2M_\phi\frac{\epsilon\cdot q}{q^2}q^\mu A_3(q^2)
-2M_\phi\frac{\epsilon\cdot q}{q^2}q^\mu A_0(q^2),
\eeqan
where
\begin{equation}
A_3(q^2)=\frac{M_{D_s}+M_\phi}{2M_\phi}A_1(q^2)-
\frac{M_{D_s}-M_\phi}{2M_\phi}A_2(q^2)\textrm{ and }A_3(0)=A_0(0).
\end{equation}
Here $p$ is the momentum of the $D_s$, $p'$ is the momentum of the
$\phi$ and $\epsilon$ is the polarisation vector. $V(q^2)$ is the
vector form factor and $A_0(q^2)$, $A_1(q^2)$, $A_2(q^2)$, $A_3(q^2)$
are axial vector form factors. Note that only three of the axial
vector form factors are independent. We can
extract the different form factors by choosing the right kinematics
-- see~\cite{Lattice2011Gordon} for more details. The diagram
on the right in Fig.~\ref{fig:spectatordiagram} shows the kinematics
of this decay: $\theta_e$ is the angle between the $e^+$ and the $W$
and $\theta_K$ is the angle between the kaon and the $\phi$. $\theta_e$
is measured in the rest frame of the $W$ and $\theta_K$ is measured
in the rest frame of the $\phi$. Our results, the form factors and
angular distributions, are plotted in Figs.~\ref{fig:DsphiFF} and
\ref{fig:Dsphi}. The `boxes' in Fig.~\ref{fig:Dsphi} show
the experimental results from BaBar~\cite{BaBar} for the angular
distributions -- we see good agreement between theory and experiment.

\subsection{Charm to light decay:
$D\to \pi\ell\nu$ and $D_s\to K\ell\nu$}

\begin{figure}[htb]
\centering
\includegraphics[width=0.75\textwidth]{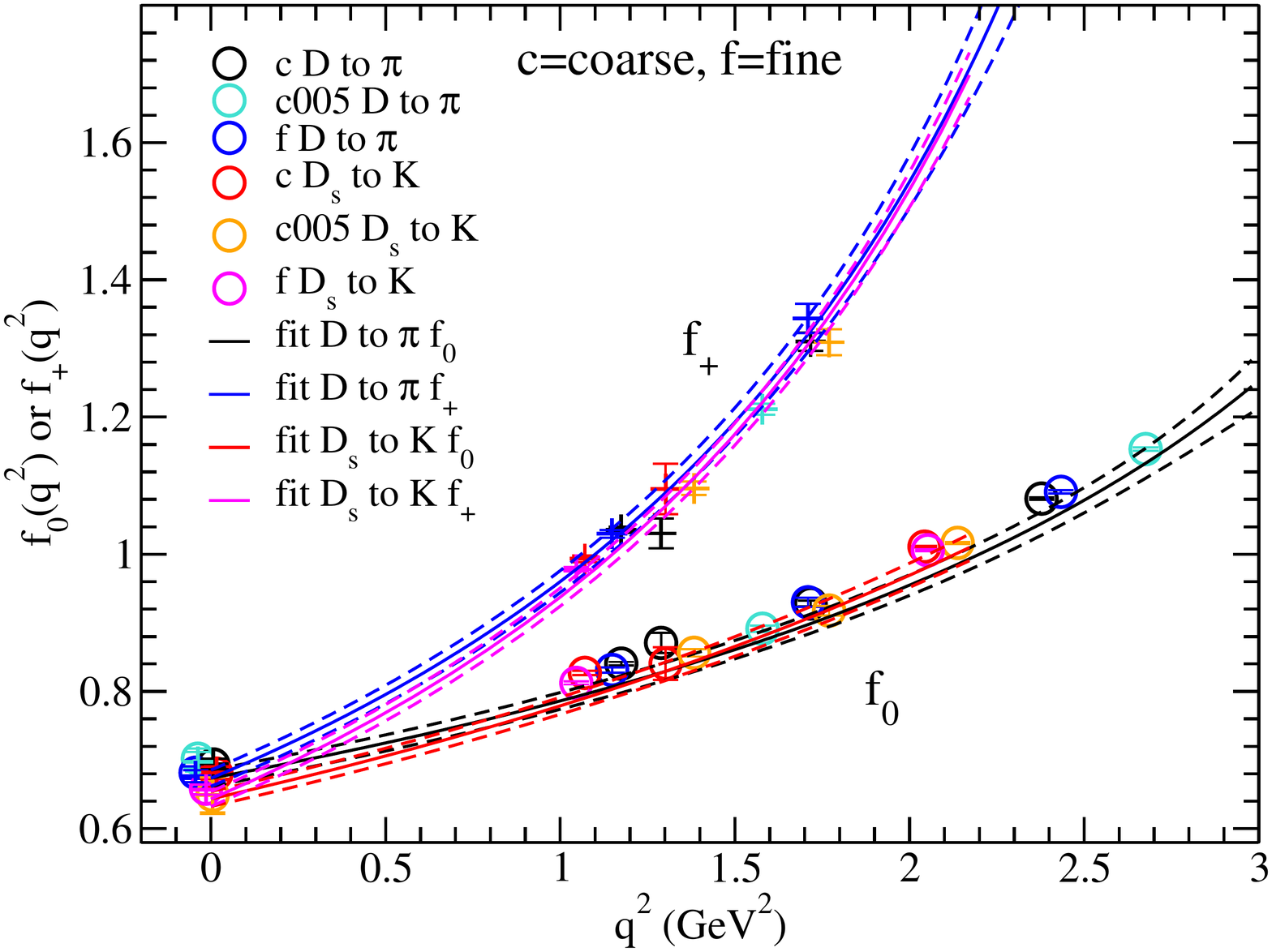}
\caption{$D\to \pi\ell\nu$ and $D_s\to K\ell\nu$ form factors.
The form factors are not sensitive to the spectator quark.}
\label{fig:DsKnDpi}
\end{figure}

We calculate two charm to light semileptonic decays, $D\to \pi\ell\nu$
and $D_s\to K\ell\nu$. In this case, both of the decays are experimentally
accessible. Our results on different lattice ensembles, along with the
final result after the continuum and chiral extrapolation
(see Section~\ref{sect:zexpansion}), are shown in Fig.~\ref{fig:DsKnDpi}.
The conclusions are very similar to the charm to strange decay: Again,
the dependence of the form factors on the spectator quark mass is very
mild -- going from a strange spectator quark to a light quark changes
the form factors by less than 5\%.

\section{The $z$-expansion}
\label{sect:zexpansion}

\begin{figure}[htb]
\begin{center}
\includegraphics[width=0.6\textwidth]{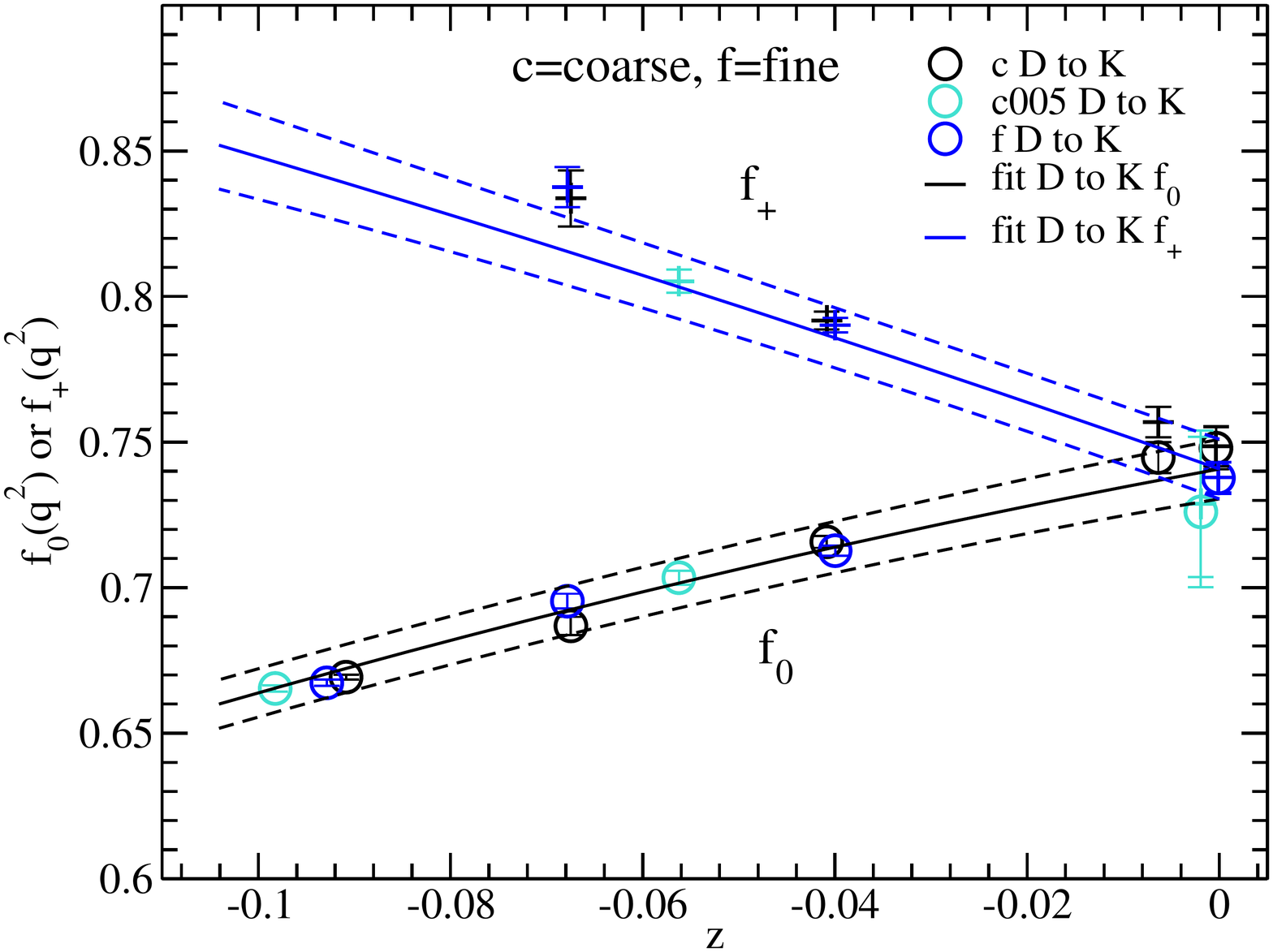}
\includegraphics[width=0.39\textwidth]{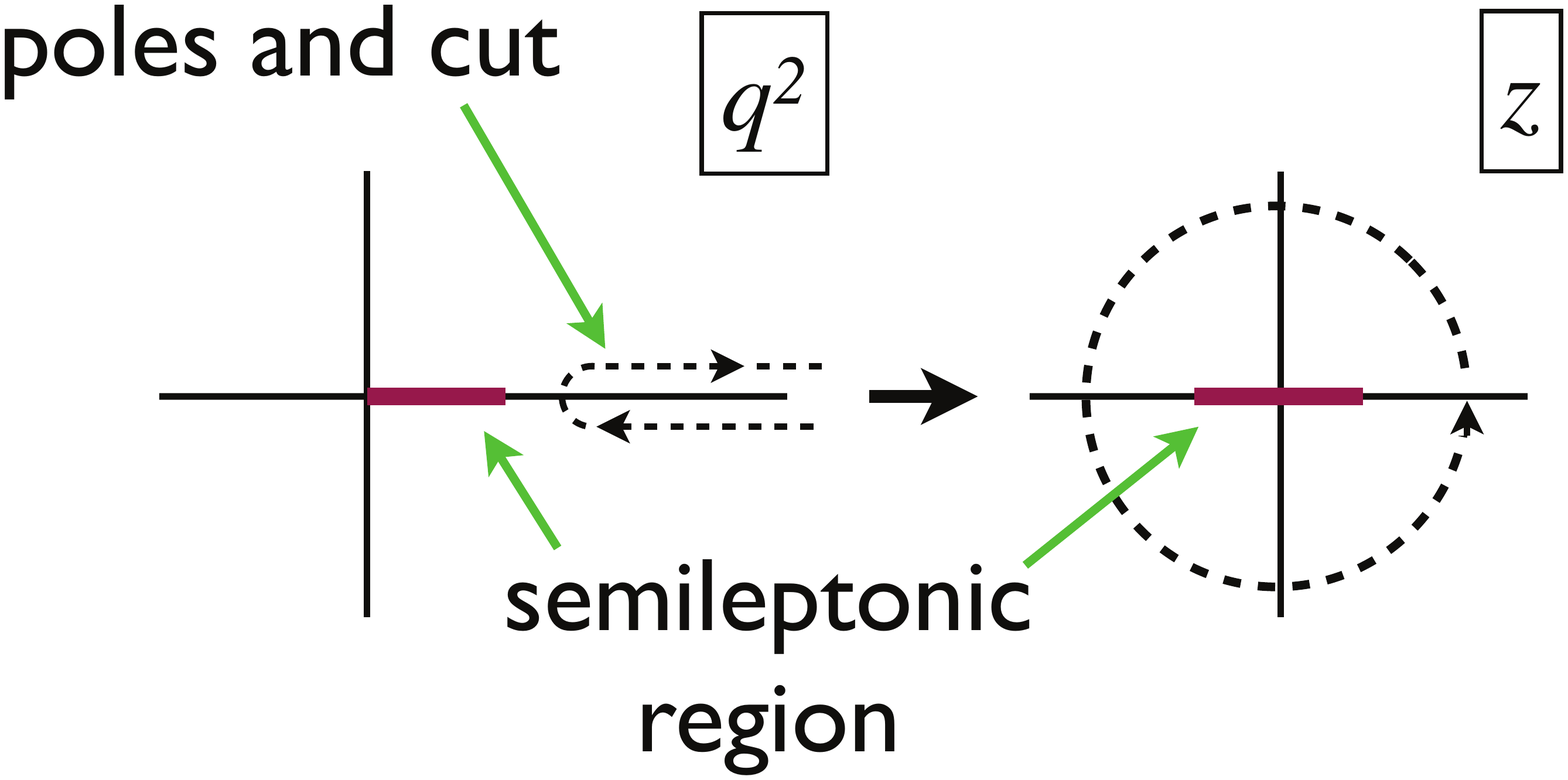}
\caption{On the left: $D\to K\ell\nu$ form factors in $z$ space
and the fit. On the right: Transformation of the complex $q^2$
plane to the $z$ plane.}
\label{fig:DKzspacefit}
\end{center}
\end{figure}

For continuum and chiral extrapolations we use the
z-expansion~\cite{zexpansion}: First we remove the poles
(using $D\to K$ decay here as an example)
\begin{equation}
\tilde{f}_0^{D\to K}(q^2)=\Bigg(1-\frac{q^2}{M_{D^\ast_{s0}}^2}\Bigg)
f_0^{D\to K}(q^2),\quad
\tilde{f}_+^{D\to K}(q^2)=\Bigg(1-\frac{q^2}{M_{D^\ast_s}^2}\Bigg)
f_+^{D\to K}(q^2)
\end{equation}
and convert to $z$ space
\begin{equation}
z(q^2)=\frac{\sqrt{t_+-q^2}-\sqrt{t_+}}{\sqrt{t_+-q^2}+\sqrt{t_+}},
\quad t_+=(M_D+M_K)^2.
\end{equation}
The transformation of the complex $q^2$ plane to the $z$ plane is
sketched in Fig.~\ref{fig:DKzspacefit}. This is useful, as
the form factors $\tilde{f}$ in the physical region can
be described by a simple power series in $z$:
\begin{equation}
\tilde{f}_0^{D\to K}(z)=\sum_{n\geq 0}b_n(a)z^n,\quad
\tilde{f}_+^{D\to K}(z)=\sum_{n\geq 0}c_n(a)z^n,\quad
c_0=b_0.
\end{equation}
We let the fit parameters $b_n$, $c_n$ depend on lattice spacing and
sea quark masses. In the end we take $a=0$ and $m_q=m_q^\mathrm{phys}$
to get the result in the physical limit. Fig.~\ref{fig:DKzspacefit}
shows one of the fits, for $D\to K\ell\nu$.

\section{Extracting $V_{cs}$}
\label{sect:extractVcs}

\begin{figure}[htb]
\centering
\includegraphics[width=0.58\textwidth]{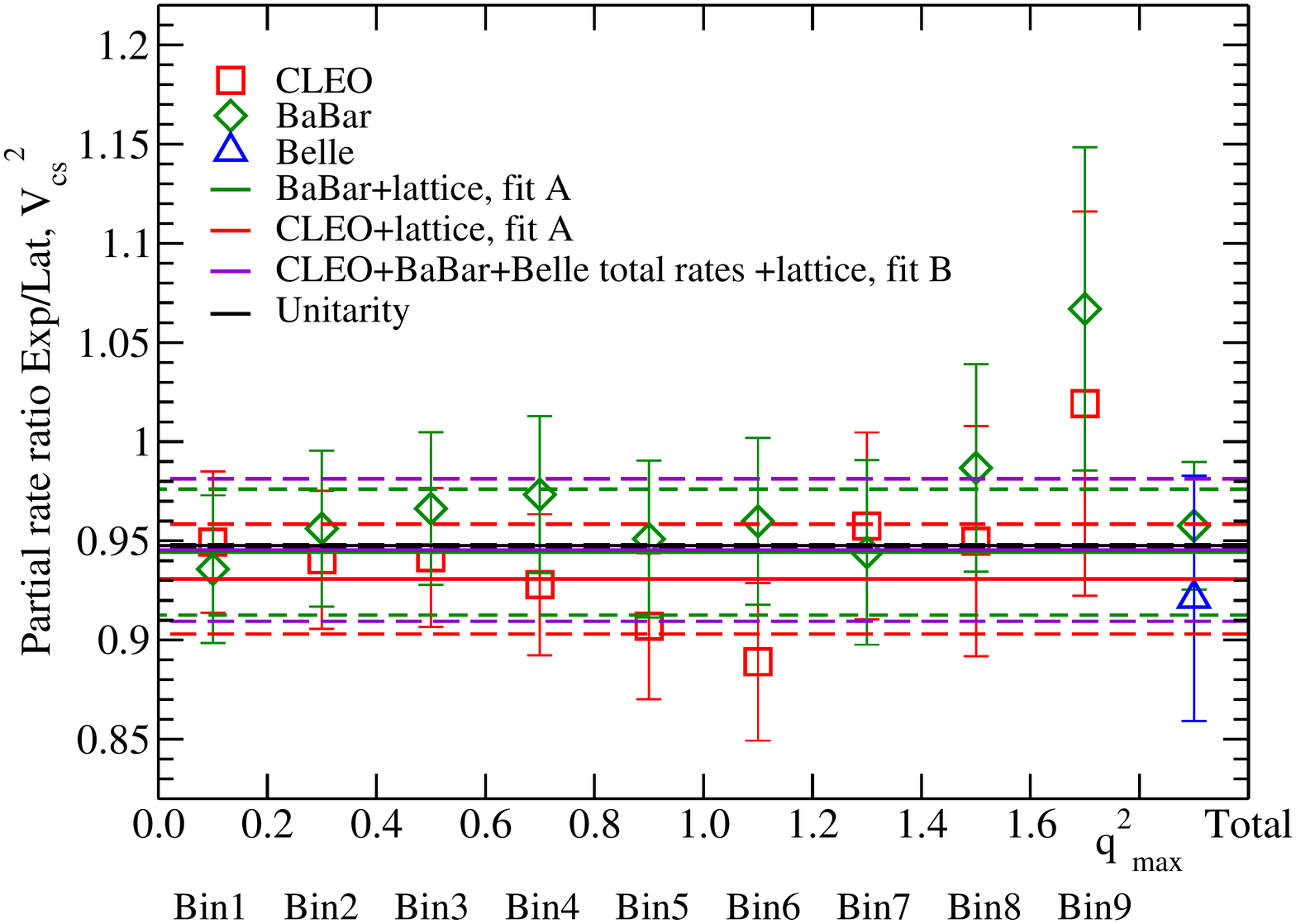}
\caption{Ratio of experimental to lattice results in each $q^2$
bin for $D\to K\ell\nu$, i.e. $|V_{cs}|^2$ extracted from that
bin directly. The experimental results are
from~\cite{CLEO, BaBar, Belle, BESIII}.}
\label{fig:explatratio}
\end{figure}

\begin{figure}[htb]
\centering
\includegraphics[width=0.73\textwidth]{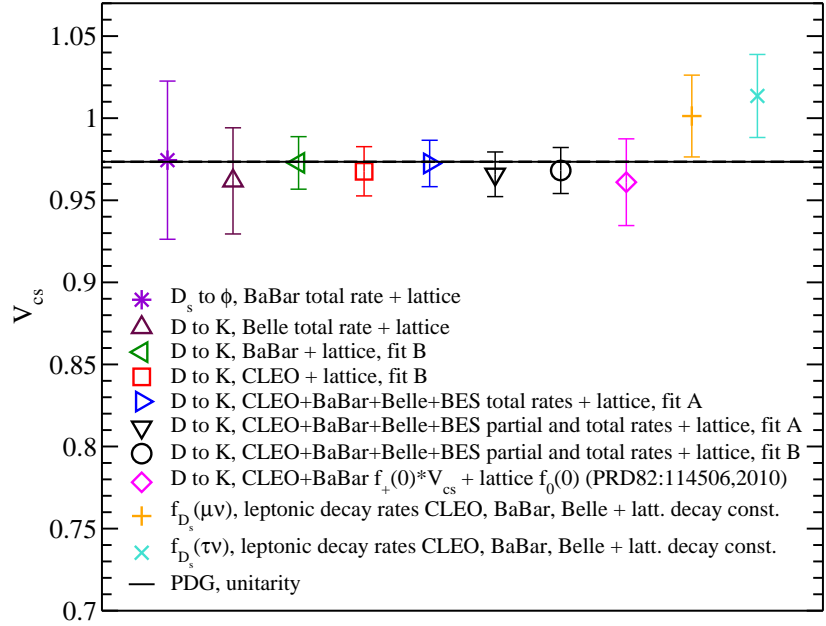}
\caption{Our determination of $V_{cs}$ from this Lattice QCD study
combined with various sets of experimental results. $D\to K\ell\nu$
experimental results are from~\cite{CLEO, BaBar, Belle, BESIII}.
The decay constant $f_{D_s}$ is from~\cite{Dsdecayconst} and
leptonic decay rates are from~\cite{leptonicdecay}.}
\label{fig:Vcs}
\end{figure}

\begin{figure}[htb]
\centering
\includegraphics[width=0.71\textwidth]{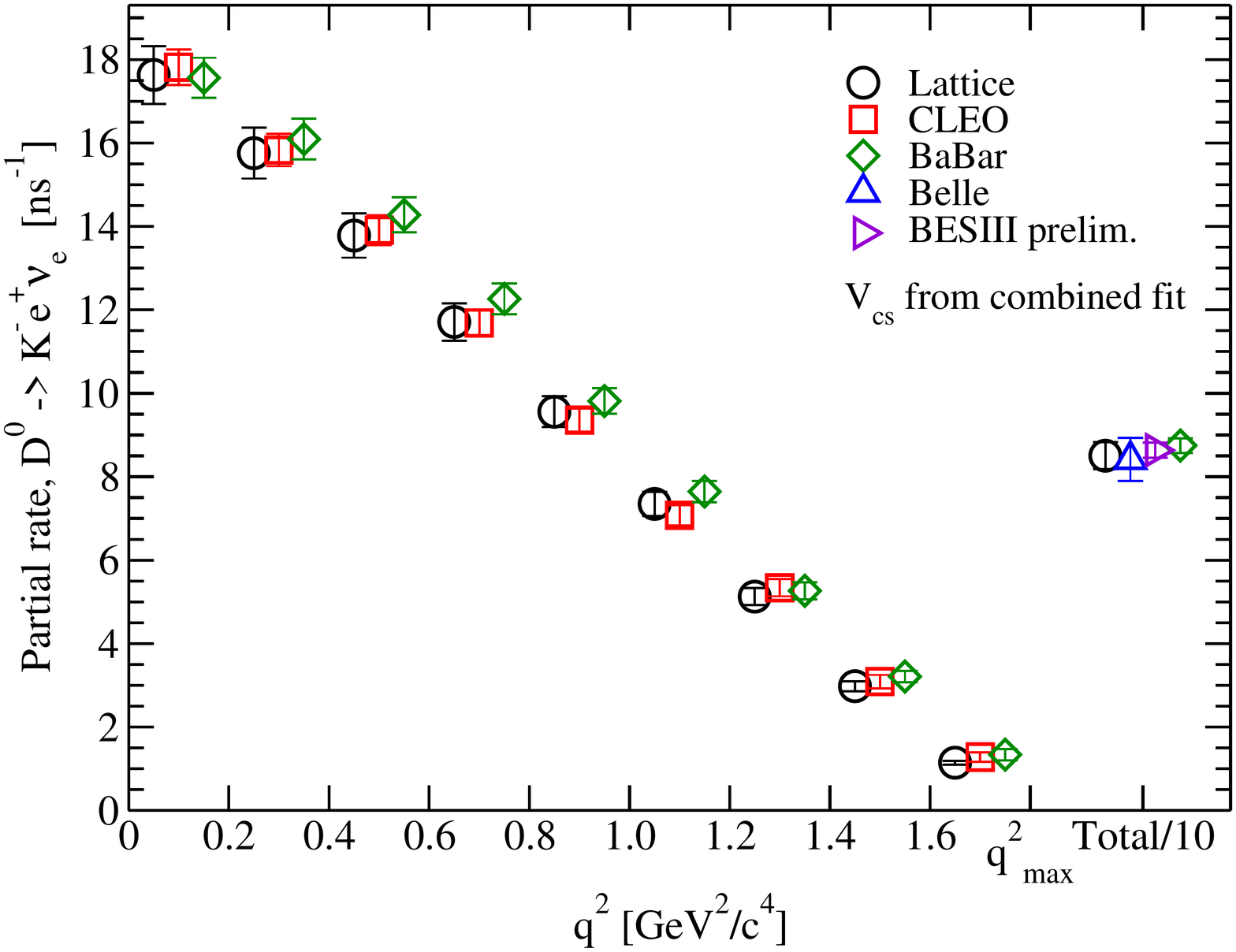}
\caption{Decay rates in $q^2$ bins for $D\to K\ell\nu$.
The lattice numbers include our result for $V_{cs}$. The
experimental results are from~\cite{CLEO, BaBar, Belle, BESIII}.}
\label{fig:rateq2binsDK}
\end{figure}

One of the goals is to determine the CKM matrix elements and
over-constrain the unitarity triangle by providing accurate input
from theory. The lattice calculation does not know about $V_{cs}$,
but experiments do. By combining the lattice calculation of the
$D\to K\ell\nu$ form factor $f_+(q^2)$ with experimental results
we can extract $V_{cs}$  without the need to assume unitarity of
the CKM matrix. By integrating the lattice form factors over the
experimental $q^2$ bins, i.e. integrating Eq.~\ref{eq:diffdecayrate}
to get the rate for a given bin and then taking the experiment
to lattice ratio gives us $V_{cs}^2$ for that given bin. Our
results are shown in Fig.~\ref{fig:explatratio}. We do this using 
CLEO~\cite{CLEO}, BaBar~\cite{BaBar}, Belle~\cite{Belle} and BESIII 
(preliminary,~\cite{BESIII}) results and fit a constant to
these  $V_{cs}^2$ values, including bin to bin correlations from 
lattice calculations and experiments in the fit. Including
different sets of experimental results does not change our value
for $V_{cs}$, as can be seen in Fig.~\ref{fig:Vcs}. Our best,
preliminary value is $V_{cs}=0.965(14)$. This is consistent with
$V_{cs} = 0.97344(16)$ from PDG~\cite{PDG}, that is calcuated
by assuming unitarity. Using our best value for $V_{cs}$ we
plot the $D\to K\ell\nu$ decay rates in $q^2$ bins in
Fig.~\ref{fig:rateq2binsDK}, which shows excellent agreement
between theory and experiments. $D_s\to\phi\ell\nu$ decay rates,
plotted in Fig.~\ref{fig:Dsphi}, also show excellent agreement.

\section{Summary}
\label{sect:summary}

This very high precision Lattice QCD calculation provides $D$
and $D_s$ meson semileptonic decay form factors from first
principles. We study the full $q^2$ range and different
semileptonic decays (different daughter mesons). We determine
the form factors to better than 3\% accuracy (better than 2\%
in the case of $D\to K\ell\nu$). The $D/D_s$ form factors are
very insensitive to the spectator quark and this is expected
to be true for $B/B_s$ as well. We calculate decay rates in
$q^2$ bins to compare with experiments and extract $V_{cs}$.
We get $V_{cs}=0.965(14)$ (preliminary), which gives very good
agreement with experiments, i.e. the shape of the form factor
$f_+^{D\to K}$ calculated in Lattice QCD agrees with experimental
results.

\Acknowledgements

We are grateful to the MILC collaboration for the use of their
configurations and to STFC for funding. Computing was done on
the Darwin supercomputer at the High Performance Computing Centre
in Cambridge as part of the DiRAC facility, jointly funded by STFC,
the Large Facilities Capital Fund of BIS and the Universities
of Cambridge and Glasgow.

\end{document}